# Tuning Electrical and Thermal Transport in AlGaN/GaN Heterostructures via Buffer Layer Engineering

*Ananth Saran Yalamarthy[1], Hongyun So[2*], Miguel Muñoz Rojo[3], Ateeq J. Suria[1], Xiaoqing Xu[4], Eric Pop[3,5,6], and Debbie G. Senesky[3,6,7*]*

[1]Department of Mechanical Engineering, Stanford University, Stanford, CA 94305, USA.
[2]Department of Mechanical Engineering, Hanyang University, Seoul 04763, South Korea.
[3]Department of Electrical Engineering, Stanford University, Stanford, CA 94305, USA.
[4]Stanford Nanofabrication Facility, Stanford University, Stanford, CA 94305, USA.
[5]Department of Materials Science and Engineering, Stanford University, Stanford, CA 94305, USA.
[6]Precourt Institute for Energy, Stanford University, Stanford, CA 94305, USA.
[7]Department of Aeronautics and Astronautics, Stanford University, Stanford, CA 94305, USA.

*E-mail: Hongyun So (hyso@hanyang.ac.kr), Debbie G. Senesky (dsenesky@stanford.edu)

Over the last decade, progress in wide bandgap, III-V materials systems based on gallium nitride (GaN) has been a major driver in the realization of high power and high frequency electronic devices. Since the highly conductive, two-dimensional electron gas (2DEG) at the AlGaN/GaN interface is based on built-in polarization fields (not doping) and is confined to very small thicknesses, its charge carriers exhibit much higher mobilities in comparison to their doped counterparts. In this study, we show that this heterostructured material also offers the unique ability to manipulate electrical transport separately from thermal transport through the examination of fully-suspended AlGaN/GaN diaphragms of varied GaN buffer layer thicknesses. Notably, we show that ~100 nm thin GaN layers can considerably impede heat flow without electrical transport degradation, and that a significant improvement (~4x) in the thermoelectric figure of merit (*zT*) over externally doped GaN is observed in 2DEG based heterostructures. We also observe state-of-the art thermoelectric power factors (4-7×10$^{-3}$ Wm$^{-1}$K$^{-2}$ at room temperature) in the 2DEG of this material system. This remarkable tuning behavior and thermoelectric enhancement, elucidated here for the first time in a polarization-based heterostructure, is achieved since the electrons are at the heterostructured interface, while the phonons are within the material system. These results highlight the potential for using the 2DEG in III-V materials for on-chip thermal sensing and energy harvesting.

## 1. Introduction

Over the past decade, gallium nitride on silicon (GaN-on-Si) substrates have gained widespread use as a materials platform for high-power[1], high-frequency[1] and extreme temperature electronics.[2] This technology is enabled by the presence of a two-dimensional electron gas (2DEG) that is formed when a nanometer-thick layer of unintentionally doped aluminum gallium nitride (AlGaN), indium gallium nitride (InGaN) or aluminum indium nitride (AlInN) is deposited on an underlying GaN buffer layer.[3–5] The 2DEG, created from built-in polarization fields and surface states in the undoped III-V layers, has a high sheet density $n_s \sim 10^{13}$ cm$^{-2}$, high room-temperature mobility $\mu \sim$ 1500-2000 cm$^2$V$^{-1}$s$^{-1}$ and has been reported to operate at temperatures as high as 1000°C.[2]

GaN thin films have been utilized in the design of thermoelectric devices because of their high-temperature operation and potential for on-chip energy harvesting and sensing.[6] As a result, the electro-thermal transport in various GaN-based materials is being investigated, including externally n-doped AlInN with a thermoelectric figure of merit $zT \sim 0.1$ at 25°C,[7] and bulk InGaN with a $zT$ as high as 0.34 at ~600°C.[8] Recently, 2DEG heterostructures[9,10] and GaN-based superlattices have attracted great attention in the design of thermoelectric devices as bulk doped materials are limited by lower mobilities[11] (200 – 400 cm$^2$V$^{-1}$s$^{-1}$). For example, recent experiments showed ~10x improvement in power factor ($S^2\sigma$) over bulk doped GaN (as high as $2\times10^{-3}$ Wm$^{-1}$K$^{-2}$ using an AlGaN/GaN superlattice[10]), as well as simultaneous increase in Seebeck coefficient ($S$) and electrical conductivity ($\sigma$) for the 2DEG,[9] contrary to bulk doped materials. The primary contributor in the power factor enhancement is the improved 2DEG mobility,[10] yet, the overall thermoelectric efficiency is limited by the high thermal conductivity[12] of its underlying GaN buffer (confinement) layer. To date, researchers have overlooked the ability to modify the GaN buffer layer to a reduced size in order to significantly scatter the phonons in it, while simultaneously preserving the electrical integrity of the 2DEG. Futhermore, the thermoelectric properties of material systems where the interfacial charge is caused by polarization fields (e.g., AlGaN/GaN) have not been investigated to date over a wide range of temperature and buffer configurations.

In this communication, we present this remarkable ability to independently manipulate electrical transport separately from thermal transport in AlGaN/GaN heterostructures. This tuning behavior is possible as it arises from polarization fields, the electrons are tightly confined at the interface, while the phonons are in the material system.[10] Our device test platform is composed of a fully-suspended, microfabricated device architecture that enables characterization of the in-plane thermal and electrical transport in AlGaN/GaN heterostructures with varying buffer layer thickness. Figure 1 shows a conceptual schematic of the aims of this study. Transport properties are studied over a wide temperature range from 25°C to 300°C. Notably, we show that ~100 nm thin GaN layers can considerably impede heat flow without significant electrical transport degradation, and that a large improvement (~4x) in the thermoelectric figure of merit over externally doped GaN is observed in 2DEG based heterostructures. Furthermore, our experiments also demonstrate state-of-the-art[13] thermoelectric power factors (4-7×$10^{-3}$ Wm$^{-1}$K$^{-2}$ at room temperature) observed in the 2DEG of this material system. This remarkable tuning behavior and thermoelectric enhancement, elucidated here for the first time in the AlGaN/GaN 2DEG heterostructure, demonstrates how tuning the polarization fields at material interfaces can be used for thermal sensing and energy harvesting applications.

**2. Test structures and measurements**

Figure 2a and Figure 2b are microscope images of our two fully-suspended heterostructure platforms for the measurement of in-plane thermal conductivity of the heterostructure stack and Seebeck coefficient of the 2DEG. A scanning electron microscope (SEM) image of a cross-section of the suspension region is shown in Figure 2c. These structures are based on the well-known central line heater method used in thermal characterization.[14,15] The suspended platform was micro-fabricated using a seven-mask process (Supporting Information Section 1), with deep reactive ion etching (DRIE) used as the final processing step to remove the Si(111) below the heterostructure, as seen in Figure 2c. Two parallel, ~5 μm wide Ti/Pt metal lines separated by 75 μm are used as heater and sensor thermometers, patterned on a ~47 nm thick amorphous $Al_2O_3$ layer that provides electrical isolation from the heterostructure. For Seebeck coefficient measurement, only a heater thermometer is patterned adjacent to a 2DEG mesa with Ohmic contacts extending to the substrate, as illustrated in Figure 2b. We used an in-house metal organic chemical vapor deposition (MOCVD) system (Aixtron, Inc.) to deposit the AlGaN/GaN/buffer heterostructure layers on top of p-type Si(111) substrates with resistivity of 0.1−1 Ω-cm. Additional details about the growth process can be found in our former work.[16] The buffer layers ($Al_xGa_{1-x}N$, $0 \leq x \leq 1$) are unintentionally doped below $10^{16}$ $cm^{-3}$. Current-voltage (I-V) measurements after etching the 2DEG mesa were below the measurement resolution of our system (~10 pA), which supports the assumption of $R_{buffer} \gg R_{2DEG}$, where $R$ is the resistance. Thus the buffer layers can be considered semi-insulating. Two variants of the heterostructure with GaN thicknesses of 1.2 μm and ~100–150 nm are grown and called the "bulk GaN" (Figure 2d) and "thin GaN" (Figure 2e) samples, respectively. The "bulk GaN" heterostructure is a still a thin film, and reflects the heterostructure thicknesses that are typical for AlGaN/GaN power devices.[17] The selection of the GaN thickness in the "thin GaN" device structure is based on a trade-off to reduce the thermal conductivity of the buffer structure while preserving the 2DEG conductivity (Supporting Information Section 1). Forming of the 2DEG was accomplished by depositing ~30 nm of unintentionally doped $Al_{0.25}Ga_{0.75}N$ barrier layer on the GaN layer in both heterostructure variants. A thin GaN capping layer of ~3 nm was grown on top of the AlGaN barrier layer, and a 1 nm thick AlN spacer was inserted between the AlGaN and the GaN layers for 2DEG mobility enhancement.[18]

Measurement of the in-plane thermal conductivity is conducted as follows. The sample is attached to a temperature controlled chuck (Signatone Inc.) via a vacuum-compatible thermal grease (Apiezon Inc.) with air as the ambient. We pass a range of DC currents through the heater metal line to induce a temperature gradient in the heterostructure (Figure 2f) and simultaneously measure the electrical resistance of the metal electrodes. Typical current values are chosen to induce a maximum $\Delta T \sim 20$ K referenced to the substrate temperature, which varies from 25°C to 300°C. The placement of the sensor electrode was carefully designed to allow for a one-dimensional (1-D) in-plane heat transfer approximation in the diaphragm.[14] The electrical resistance of the electrodes was calibrated over the entire temperature range using sufficiently low currents to avoid self-heating (Supporting Information Section 2). The calibration allows us to convert the electrical resistance into corresponding temperature values using the measured temperature coefficient of resistance (TCR). From the collected temperature data, we can infer the in-plane thermal conductivity of the heterostructure given the heater power ($P_H$), after accounting for errors due to heat spreading into the $Al_2O_3$ and external losses (Supporting Information Section 3) through a simple analytical model in conjunction with a 3-D finite-element simulation. In the model, we also included estimated values of the thermal contact resistance between the electrode, insulation and heterostructure interfaces. Overall, the errors due the insulation are found to be less than ~6%, while errors due to external convective and radiative losses progressively increase to ~10% at a substrate temperature of 300°C (Supporting Information Section 3).

The measurement of the Seebeck coefficient follows a similar procedure; a current passed through the heater electrode induces a temperature gradient in the diaphragm, resulting in a Seebeck voltage across the 2DEG mesa that spans the suspension and the substrate regions (Figure 2b). Using a similar calibration procedure for the heater line, the temperature drop across the mesa can be used to extract the Seebeck coefficient, after accounting for external losses, Ohmic contact voltage drop and a minor temperature drop in the substrate (Supporting

Information Section 3). Note that the measured Seebeck coefficient corresponds to the 2DEG contribution exclusively since the III-V buffer layers are semi-insulating. Lastly, electrical conductivity of the 2DEG for the bulk and thin GaN samples are estimated using circular transfer length method (CTLM) structures with varying channel lengths ($d$ = 20 μm to 70 μm), with the aid of simulations to obtain the thickness of the 2DEG triangular potential ($t_{2D}$) well, to be discussed in the next section.

### 3. Charge Profiles

The thickness and charge density of the 2DEG for the bulk and thin GaN heterostructures are simulated using a commercially-available Schrödinger-Poisson solver (NextNano GmbH[19]). The simulated band structures and volumetric charge density profiles for the bulk and thin GaN heterostructures are illustrated in Figure 3. All the heterostructure layers are assumed to be undoped and the barrier height for the GaN capping layer is set to 1 eV.[20] For the bulk and thin GaN models, the GaN thickness and the $Al_{0.2}Ga_{0.8}N$ layer thickness (first buffer layer below the GaN) were varied until 2DEG sheet density ($n_s$) convergence was observed. In both cases, the 2DEG region is visible as a triangular potential well near the AlGaN/AlN/GaN interface. From the simulation, we found $n_s = 1.06 \times 10^{13}$ cm$^{-2}$ and $n_s = 0.91 \times 10^{13}$ cm$^{-2}$ for the bulk and thin GaN heterostructures, respectively. It should be noted that a good match, within ~10% of the theoretically calculated values is observed when comparing these values with experimental data extracted from Hall-effect devices fabricated on the same platform, which supports the model (Supporting Information Section 4). The physical thickness of the 2DEG region, $t_{2D}$, can be extracted as the region where GaN is degenerate.[9] From simulation, these thickness values were obtained to be ~6.1 nm and ~4.4 nm for the bulk and thin GaN heterostructures, respectively, which can be used to obtain the 2DEG conductivity $\sigma$ from the sheet resistance ($R_{sh}$) extracted via CTLM measurements. Finally, we note that an average 2DEG volumetric density can be estimated as $n_v = n_s/t_{2D}$ for the bulk GaN (1.73 × 10$^{19}$

cm$^{-3}$) and thin GaN (2.07 × 10$^{19}$ cm$^{-3}$) heterostructure. We note that the higher $n_v$ for the thin GaN sample reflects the smaller 2DEG quantum well thickness.

### 4. Electrical and thermal property measurements

The measurements of $R_{sh}$ averaged over four samples up to 300°C via CTLM measurements (Supporting Information Section 5) can be combined with the 2DEG thickness $t_{2D}$ to obtain the average electrical conductivity [σ = 1/($R_{sh}$×$t_{2D}$)] of the electrons in the 2DEG. At room temperature, we obtained $R_{sh}$ values of ~350 Ω/sq and ~500 Ω/sq for the bulk and thin GaN samples, respectively. We note that these values are amongst the lowest reported $R_{sh}$ values for AlGaN/GaN 2DEGs, which highlights the quality of our samples.[16] The average conductivity in the thin GaN sample is observed to be similar to the bulk GaN sample due to simultaneous reduction in the sheet density and quantum well depth, as seen in Figure 4a. Sheet densities in this temperature range are approximately constant due to negligible strain relaxation in the heterostructure layers,[37] stable piezoelectric coefficients,[22] and minimal intrinsic carrier concentration change due to the wide-bandgap. Thus, the decrease of $\sigma$ at high temperatures is mainly determined by the 2DEG mobility, $\mu$. The dependence is well described by a temperature power law ~$T^{-2.5}$ that arises from electron-optical phonon scattering, which further supports this fact. We also note that our exponent is consistent with former exponent ranges (from -2.2 to -3.4) reported in the literature.[21]

Temperature dependent in-plane thermal conductivity measurements for the bulk and thin GaN samples are shown in Figure 4b. Room temperature thermal conductivity dropped from ~115 Wm$^{-1}$K$^{-1}$ for the bulk GaN sample to ~45 Wm$^{-1}$K$^{-1}$ for the thin GaN sample due to phonon boundary scattering, i.e. the size effect.[15] The measurements for the bulk GaN sample follow a $T^{-1.18}$ fit. This is consistent with a similar temperature exponent observed in measurements of cross-plane thermal conductivity measurements of GaN films of thickness of ~0.7 μm.[12,23] However, for the thin GaN sample, we note that the measured thermal conductivity values follow a $T^{-0.88}$ fit, indicating that Umklapp scattering is less prominent for long wavelength phonons which are suppressed due to size effect.

In-plane thermal conductivity data in these films are limited, with little data available on the size effect and temperature dependence.[12] Since our suspended film is a composite consisting of an AlN layer, $Al_xGa_{1-x}N$ transition layers and a GaN layer, the overall thermal conductivity ($k$) can be estimated as $\sum k_i t_i / \sum t_i$, where $k_i$ and $t_i$ refer to the thermal conductivities and thicknesses of individual layers. For each multilayer, we used a Boltzmann Transport Equation (BTE) model to quantify $k_i$ with layer thickness ($t_i$). Using a simple Debye approximation for the phonon dispersion with an average velocity over the acoustic phonon modes ($v_{ac}$), the in-plane thermal conductivity for each layer can be written as[24]

$$k_i = \frac{3k_B^4 T^3}{8\pi^3 \hbar^3 v_{ac}^3} \int_0^{\theta_D/T} \int_0^{2\pi} \int_0^{\pi} \sin(\theta)\, \tau_C(x) x^2 v_g^2 d\theta d\phi dx \qquad (1)$$

where $k_B$ is the Boltzmann constant, $\theta_D$ is the Debye temperature for the multilayer,[25] $T$ is the temperature, $\hbar$ is the reduced Planck's constant, and $x = \hbar\omega/k_B T$. Here, $\omega$ is the phonon frequency. The integration is performed over the angular directions ($\theta$ and $\varphi$) using a direction dependent group velocity $v_g = v_{ac}\sin(\theta)\cos(\varphi)$.

The total scattering time $\tau_C$ is calculated by Mathiessen's rule with contributions from Umklapp ($\tau_U$), impurity ($\tau_I$), alloy ($\tau_A$), boundary ($\tau_B$) and defect scattering ($\tau_D$), respectively. The Umklapp scattering term is evaluated via the Callaway relationship, $\tau_U = A/\omega^2$. We evaluated constant $A$ in the bulk limit as $2\pi^2 v_{ac} k_\infty / k_B \omega_D$, where $k_\infty$ in the bulk thermal conductivity of layer and $\omega_D$ is the Debye frequency. For instance, $k_\infty$ values of 240 Wm$^{-1}$K$^{-1}$ and 285 Wm$^{-1}$K$^{-1}$ are used for GaN and AlN at room temperature, respectively.[26,27] The Debye frequencies are extracted from the known Debye temperatures of these material layers.[25] Scattering with impurities is neglected since its effect is found to be negligible for unintentionally doped films.[26] For the underlying transition layers, all the material parameters (e.g. $v_{ac}$, $\theta_D$, $k_\infty$) are averaged over the AlN and GaN fractions, in context of the virtual crystal model.[28] Alloy scattering severely reduces the thermal conductivity of the transition layers and is evaluated as a point defect scattering term.[29] For the sake of brevity, we skip the details, which can be found in Liu *et al.*[29] The defect scattering term ($\tau_D$) included core, screw, edge and mixed dislocations with total density $N_{dis}$, whose effect is to reduce the thermal conductivity.[30]

Although we have a composite film (and thus, the dislocation density is expected to vary for the different layers), we estimated an average value for the composite film via X-Ray Diffraction (XRD) measurements. For example, Figure 4c shows symmetric (rocking curve) scans of the (0002) lattice planes in the thin and thick GaN samples. Using the full width at half maximum (FWHM, denoted by β) value of the XRD scans, the screw-type dislocation density can be estimated as $\approx \beta^2/(4.35c^2)$, where c is the lattice constant (0.5185 nm) along the c-axis direction for the GaN wurtzite crystal.[17] Using this, we calculate the screw-type dislocation density to be $\approx 9\times10^8$ cm$^{-2}$ and $\approx 2.5\times10^9$ cm$^{-2}$ for the thick and thin GaN samples, respectively. An estimate of the edge-type dislocation density from XRD measurements is shown in Supporting Information Figure S10a, also in the order of $10^9$ cm$^{-2}$. Finally, the direction-dependent boundary scattering term[24] is evaluated as $\tau_B = t_i/2v_{ac}|\cos(\theta)|$. The modeled in-plane thermal conductivities of the individual layers in the composite stack are depicted in Figure 4d as a function of the corresponding layer thickness.

Due to the lack on in-plane thermal conductivity data in these films, it is difficult to compare the model in Figure 4d with literature. However, an estimate of the cross-plane thermal conductivity[24] using the layers can be estimated using a different $v_g = v_{ac}\cos(\theta)$ in Equation 1, the results of which are in Supporting Information Figure S10b, showing good agreement with previous measurements. Using the thicknesses of the multilayers (Figure 2), the overall in-plane thermal conductivity of the composite stack (*k*) where the GaN buffer layer thickness ($t_{GaN}$) is varied (at room temperature) is shown in Figure 4e. In particular, we note that the model agrees with the experimental data well, and that dislocation densities in our range ($10^9 - 10^{10}$ cm$^{-2}$) are estimated to have little effect on thermal conductivity of the film layers. In other words, the in-plane thermal conductivity reduction from the thick GaN to the thin GaN composite stack is expected to arise almost exclusively from the size effect. We note that this observation is consistent with former reports of the cross-plane thermal conductivity in GaN films with similar dislocation densities.[30,31]

The measured Seebeck coefficients for the bulk and thin GaN heterostructures are shown in Figure 4f.[25] While the Seebeck coefficients for bulk doped III-V films have been reported in the past,[8,11,25] values for a polarization induced 2DEG have not been well studied in literature. The approximate linear increase in the Seebeck coefficient with temperature indicates a degenerate semiconductor that may be well approximated with a Cutler-Mott[39] formula for nearly free electrons. Since $R_{buffer} \gg R_{2DEG}$, the measured values can be considered to arise exclusively from the 2DEG,[33] which is in agreement the degenerate semiconductor characteristic. To understand the magnitude and the observed dependencies with temperature, we implemented a simple analytical model based on Sztein *et al.*[25] For a bulk doped material with fixed electron concentration, the Seebeck coefficient is analytically evaluated as[25] evaluated as $S = -(\int(E-E_F)\sigma(E)dE)/(qT\int\sigma(E)dE)$, where $\sigma(E)$ is the differential electronic conductivity, $q$ is the fundamental charge, $E$ is the electron energy and $E_F$ is the Fermi energy.

Following Sztein *et al.*,[25] we evaluated $S$ as a function of doping densities in GaN. The model for the Seebeck coefficient shows good match with measurements for doped GaN films (Supporting Information Figure S11). These doping densities only serve to "mimic" the effect of 2DEG charge density, since the origin of the 2DEG is related to built-in polarization fields as discussed earlier. The differential electronic conductivity is evaluated using a non-parabolic energy dispersion relation and typical scattering mechanisms found in III-nitrides: optical phonon, piezoelectric, deformation potential and charged dislocation scattering.[25] The parameters and assumptions in the model follow from Sztein *et al.*,[25] with the notable exception that we neglected ionized impurity scattering in the evaluation of $\sigma(E)$, since the 2DEG region is assumed to be undoped. In principle, the Seebeck coefficient for the 2DEG region can be evaluated by discretizing it into fine regions with approximately constant charge concentration using Equation 2 via a thickness average. For simplicity, we instead compare the average volumetric density $n_v$ for the 2DEG region that we obtained by calculating the average charge density $n_s/t_{2D}$ against the simulated Seebeck coefficients for "mimic" doping densities. The predicted Seebeck coefficient dependence using $n_v$ (= 1.73 × 10$^{19}$ cm$^{-3}$) is in good agreement with the measured values for the 2DEG for the bulk GaN sample, as illustrated in Figure 4f. However, the Seebeck coefficient values for the thin GaN sample are lower than the bulk GaN sample, which could arise from a larger $n_v$ (= 2.07 × 10$^{19}$ cm$^{-3}$) due to reduced 2DEG thickness and differing relevance of scattering mechanisms (e.g., greater dislocation scattering) in comparison the bulk GaN film.

Finally, we also note that our measured values for the AlGaN/GaN 2DEG Seebeck coefficients do not show a similar enhancement in comparison to a bulk 3D Seebeck coefficient, which is unlike other systems, such as ZnO based 2DEG.[34] This could arise from the relatively large 2DEG confinement thickness ($t_{2D}$) of ~5 nm in our structures, as opposed to much smaller confinement depths (~1 nm) in the ZnO based 2DEG.[34] Nonetheless, our results show that tuning the thermal transport while maintaining the electrical transport via buffer layer engineering can be achieved and is useful for the design of AlGaN/GaN devices.

5. Applications

The measured electrical, thermal, and thermoelectric properties could be important for a variety of GaN-based temperature, power electronics, heat-flux, magnetic field and energy harvesting devices using III-V heterostructures. As an example, we demonstrate how the measured electrical and thermal properties could be used to detect local temperatures from on-chip heat loads using the fully-suspended bulk GaN platform. Figure 5a shows an on-chip circuit with four resistive heat sources (similar to the heater and sensor metal electrodes) with six 2DEG mesa regions (spanning the suspension and substrate regions) across which the Seebeck voltage can be measured. The substrate is held at 25°C using a temperature controlled probe-station. We then apply a heating power (25 mW) to induce a temperature gradient relative to the substrate in the suspended heterostructures. The simulated temperature profile for a quarter region using the measured thermal properties and external losses (Supporting Information Section 3) is shown in Figure 5b. Three sample locations of the "hot" ends of the 2DEG mesa where temperature is measured are illustrated in Figure 5b. Finally, with the heater power held constant, we measured the differential Seebeck voltage for the six 2DEG mesa regions. Using the measured temperature dependent Seebeck voltage for the bulk GaN film, we extracted the temperatures at the hot ends of the 2DEG mesa regions and compared them with the simulated local temperatures from the finite element model. Excellent agreement is observed (within a maximum of ~ 15% error), as seen in Figure 5c, which demonstrates how the measured properties could be used for local on-chip monitoring of thermal loads in addition to validating our measurements.

Finally, Figure 5d shows the temperature dependent thermoelectric figure of merit ($zT$) up to ~300°C using the measurements of 2DEG electrical conductivity, Seebeck coefficient and in-plane heterostructure thermal conductivity. These $zT$ values are 3 to 4x higher than in bulk doped GaN films,[10] owing to large 2DEG conductivities and thermal conductivity reduction from size effect. Further, we notice that the thin GaN film displays almost ~22-51% higher $zT$ values in comparison to the bulk GaN film due to the large, ~2.5x decrease in the in-plane thermal conductivity. The trend with temperature is almost linear, which is different from the super-linear trend[25] observed for doped III-V films, due a greater temperature exponent for the 2DEG mobility. Since the background doping concentration is low (<$10^{16}$ cm$^{-3}$) and the layers used are wide band gap materials, it is expected that the thermopower peak occurs at much higher temperatures[25] in comparison to doped III-V films, which warrants further investigation of these properties beyond 300°C. Apart from the thermoelectric figure of merit, the 2D electron gases in these films also show very high power factors between 4-7×$10^{-3}$ Wm$^{-1}$K$^{-2}$ at room temperature (Supporting Information Figure S12), which is comparable to the values for state-of-the-art thermoelectric materials.[13] These observations suggest that this approach of engineering the heterostructure layers in conjunction with polarization-based 2DEGs has promising potential for thermoelectric applications at high temperatures.

## 6. Conclusion

In summary, we designed and implemented a suspended AlGaN/GaN heterostructure platform to investigate temperature dependent thermal and electrical transport of by tuning the GaN layer thickness. We demonstrate effective manipulation of these properties, and in the process shed light on several transport parameters that have not been previously explored in detail in former literature. Notably, we show that thin GaN layers of ~100 nm significantly impede heat flow, but preserve the 2DEG conductivity, which could be useful for a range of GaN-based devices. We also show a simple example of how the measured properties can be used to monitor local heat fluxes on an AlGaN/GaN power device. We used a single AlGaN barrier layer in this study, which could serve as a backbone for extension to multiple, alternating GaN/AlGaN superlattice layers. Further work along these lines could include how changing the AlGaN thickness affects transport, improving the temperature dependent mobility degradation in AlGaN/GaN 2DEGs, exploring transport at higher temperatures beyond 300 °C, and exploring thermoelectric phenomena in other III-V heterostructure families with 2DEGs using this experimental platform.

**Acknowledgements**

This work was supported in part by the National Science Foundation (NSF) Engineering Research Center for Power Optimization of Electro Thermal Systems (POETS) under Grant EEC-1449548, by the NSF DMREF grant 1534279, and by the research fund of Hanyang University (HY-2017). The MOCVD experiments were conducted at the MOCVD Lab of the Stanford Nanofabrication Facility (SNF), which is partly supported by the NSF as part of the National Nanotechnology Coordinated Infrastructure (NNCI) under award ECCS-1542152. The authors would like to thank Catilin Chapin, Hannah Alpert and Ms. Karen Dowling for assistance with fabrication. We also thank Karen Dowling and Ms. Hannah Alpert for assistance with Hall measurements, and Thomas Heuser for assistance with the XRD measurements. The authors also acknowledge Prof. Andrew Alleyne and Pamela Tannous for useful discussions.

**Figures**

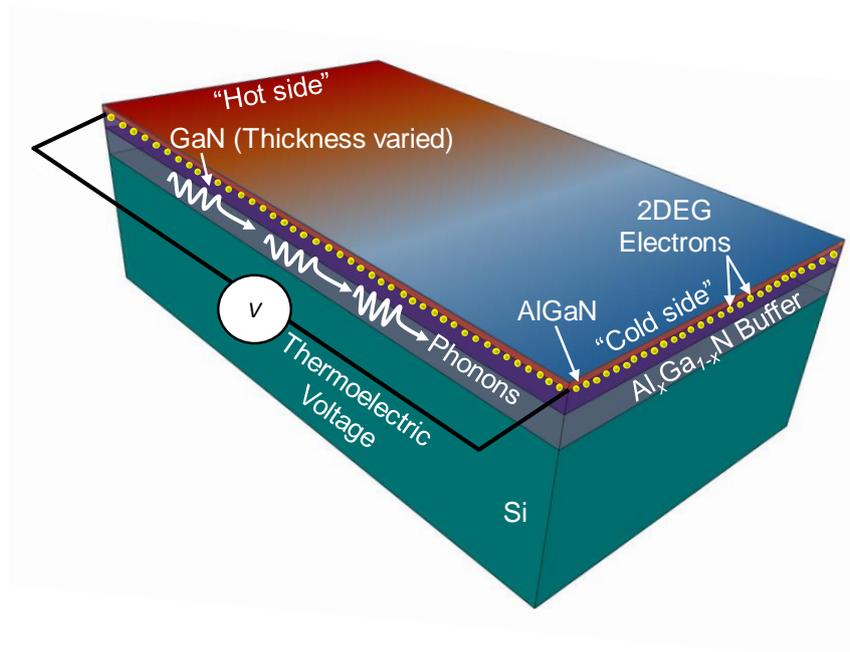

**Figure 1.** Conceptual schematic showing thermoelectric voltage generation via a lateral temperature gradient in the AlGaN/GaN 2DEG. The high mobility electrons in the 2DEG can lead to high thermoelectric power factors across the 2DEG. In this study, we explore how the thickness of the underlying GaN and buffer layers can be designed to preserve the thermoelectric power factor of the 2DEG but significantly reduce the in-plane thermal conductivity. This allows for a large improvement in the thermoelectric figure of merit in comparison with bulk doped GaN. Note that in our case we use the in-plane thermal conductivity of the GaN and the buffer (but not Si), since only these are necessary for 2DEG formation.

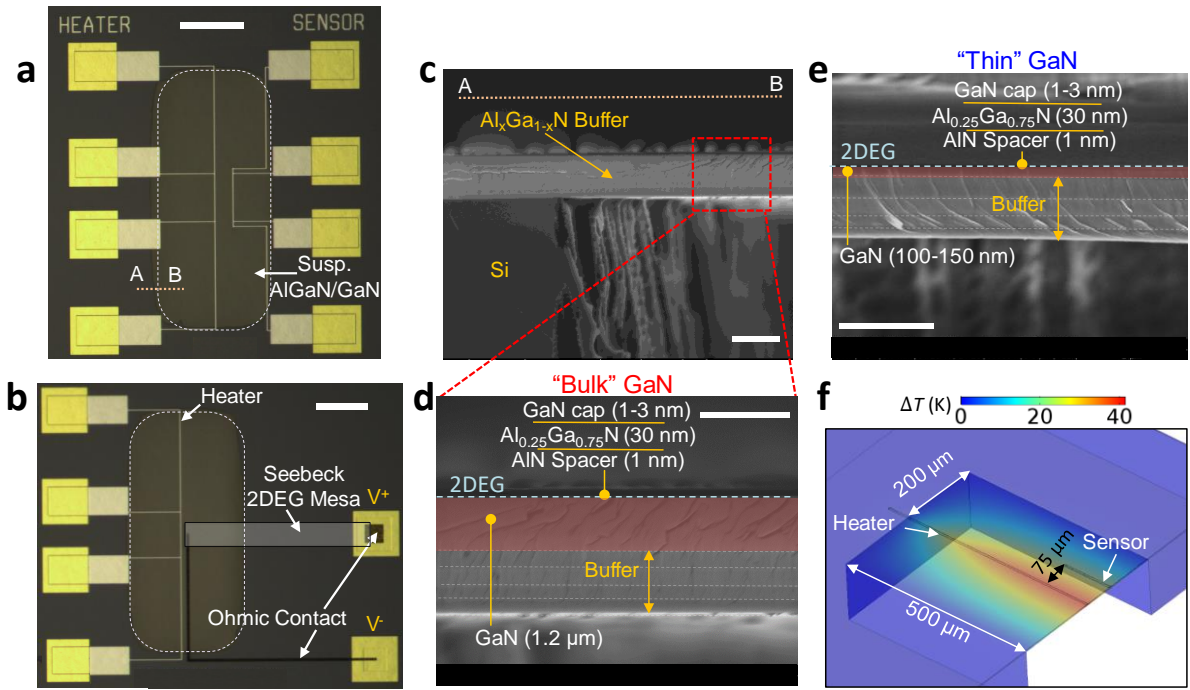

**Figure 2. (a)** Microscope image of the fully-suspended AlGaN/GaN heterostructure used for in-plane thermal conductivity measurements. The separation between the heater and sensor lines is ~75 μm. A thin ~47 nm ALD $Al_2O_3$ layer provides electrical isolation between the metal lines and the AlGaN/GaN heterostructure underneath. **(b)** Microscope image of the suspended AlGaN/GaN heterostructure to measure the Seebeck coefficient of the 2DEG. The 2DEG mesa is contacted via Ohmic Ti/Al/Pt/Au contacts. The Seebeck 2DEG mesa is shaded white for clarity. **(c)** Cross-section SEM image (of the A-B section in Figure 2a) of the suspended heterostructure, with the Si substrate selectively etched out from the backside via DRIE. **(d)** SEM image of the bulk GaN structure. The GaN thickness is ~1.2 μm and false colored. The buffer structure, starting from the Si interface is composed of AlN (300 nm)/$Al_{0.8}Ga_{0.2}N$ (300 nm)/$Al_{0.5}Ga_{0.5}N$ (400 nm)/$Al_{0.2}Ga_{0.8}N$ (500 nm). Further details are in Supporting Information Section 1. **(e)** SEM image of the thin GaN structure, showing the 2DEG at the AlGaN/GaN interface and the buffer layers. The GaN thickness is ~100-150 nm and is false colored. The other buffer layers are identical to the bulk GaN structure. **(f)** Half-symmetric finite-element simulation of experimental structure, showing sample temperature profile when current is applied through the heater (Supporting Information Section 3). **Scale bars of (a)-(e) 200 μm, 200 μm, 3 μm, 2 μm, 2 μm.**

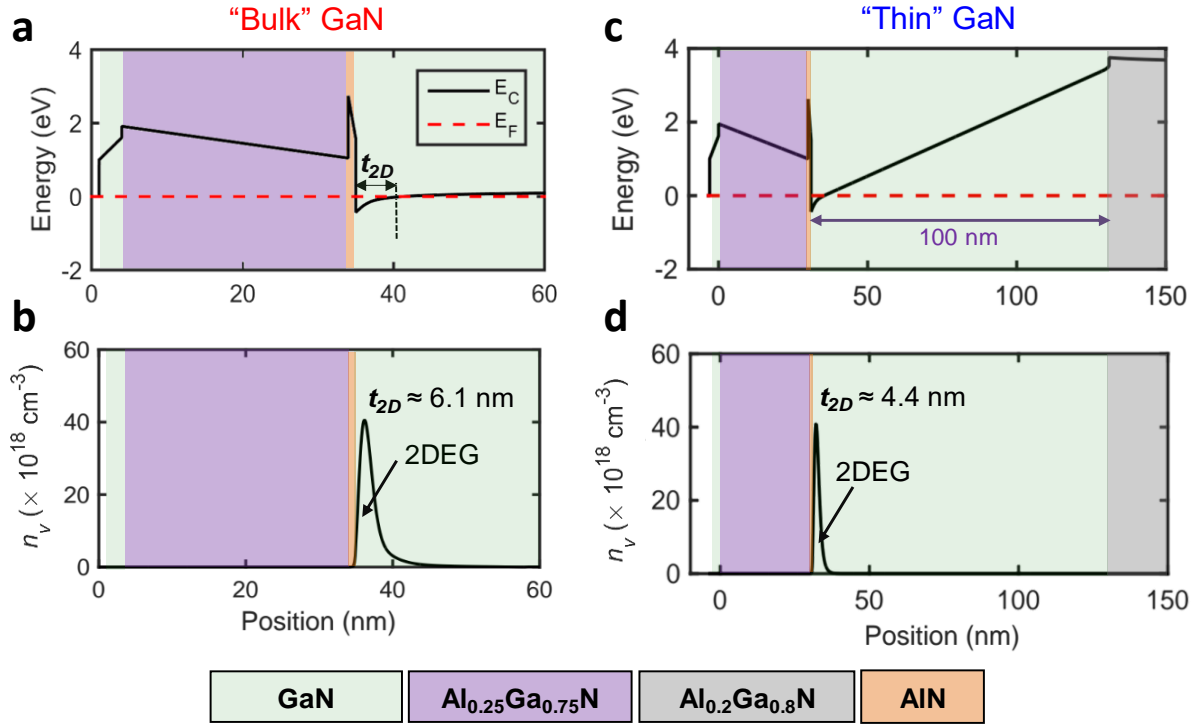

**Figure 3.** (a) Schrödinger-Poisson model of the energy band diagram for the bulk GaN structure with AlGaN thickness is 30 nm. The thickness of the 2DEG region, $t_{2D}$, is shown in the region where GaN is degenerate. (b) Volumetric charge density, $n_v$, versus position, depicting the approximately triangular charge profile, with $t_{2D} \approx 6.1$ nm. (c) Simulated energy band diagram for the thin GaN structure with AlGaN thickness of 30 nm. Note that the GaN layer is 100 nm. (d) Volumetric charge density, $n_v$, versus position, depicting the approximately triangular charge profile, with $t_{2D} \approx 4.4$ nm for the thin GaN structure.

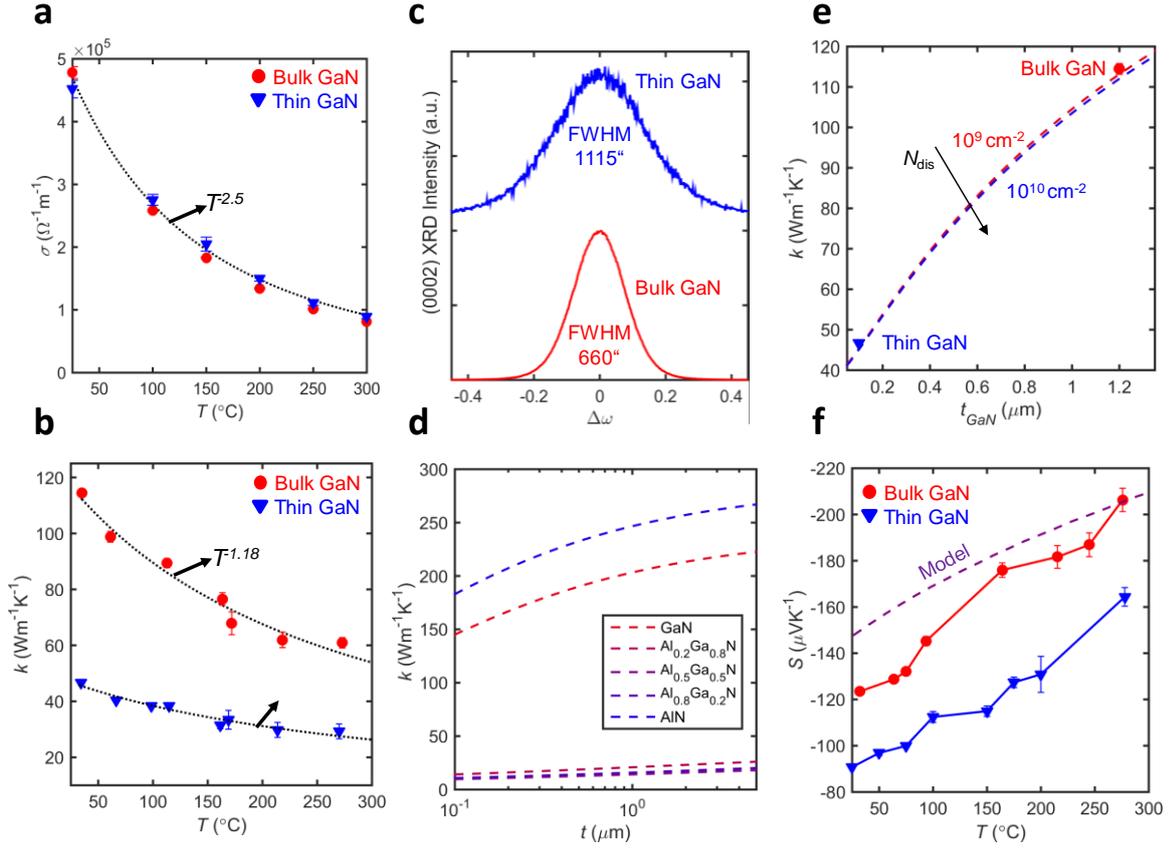

**Figure 4.** (a) Temperature dependent electrical conductivity of the 2DEG for the thin and bulk GaN samples, extracted via CTLM measurements (Supporting Information Section 5). (b) Temperature dependent thermal conductivity measurements for the bulk and thin GaN samples. Fits are shown with black, dotted lines. (c) XRD rocking curve scan of the (0002) lattice plane in GaN, to investigate the density of screw-type dislocations. (d) Modeled in-plane thermal conductivities of the layers in our composite stack as a function of thickness. (e) Thermal conductivity reduction due to size effect at 25°C. The model corresponds to the dashed lines with increasing dislocation densities ($N_{dis}$) and the data points are the experimental measurements. (f) Measured Seebeck coefficient versus temperature for thin and bulk GaN samples. The model uses $n_v \approx 2\times10^{19}$ cm$^{-3}$, which is estimated from the Schrödinger-Poisson simulation.

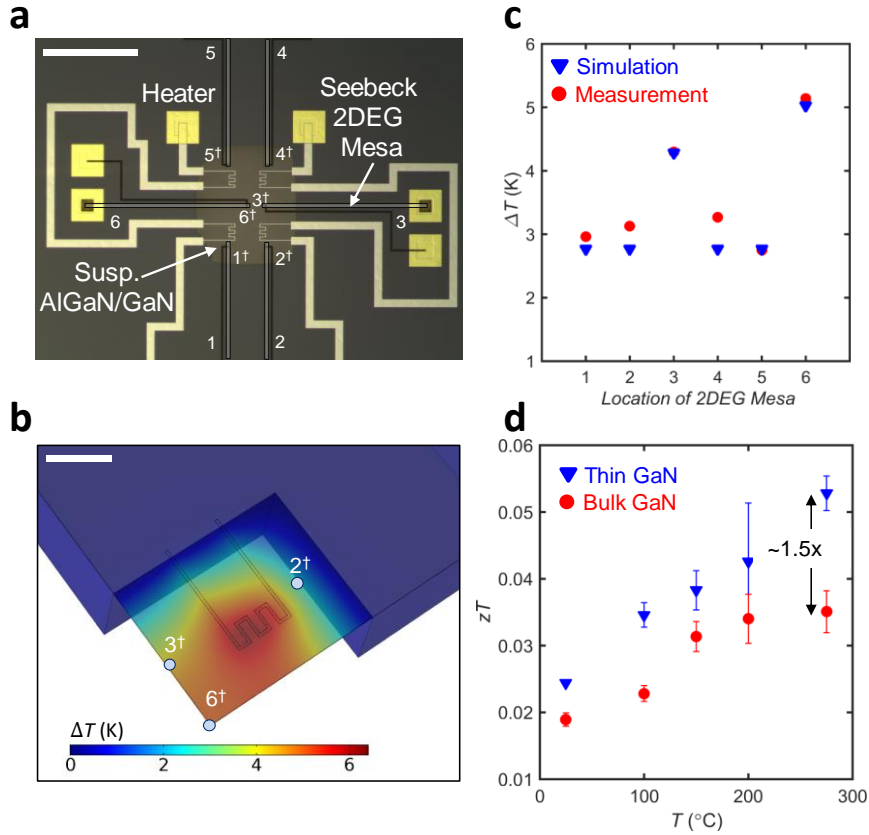

**Figure 5.** (**a**) Fully-suspended platform to measure local temperature rise using the bulk GaN heterostructure. The six labelled AlGaN-GaN Seebeck 2DEG mesa regions (e.g., 2-2†, shaded white for clarity) can be used to estimate temperature rise near the heater line, which mimics a power device. Note that the temperatures are measured at the "hot" ends of the mesa regions, marked by a † symbol. (**b**) Finite element model of temperature rise with a heater power of 25 mW using the extracted thermal properties of the bulk GaN sample. The model is for a quarter symmetry region of Figure 5a. Three locations for temperature rise measurement (2†, 3†, 6†) are also shown. The Si substrate is assumed to be at 25°C in this model. (**c**) Simulated and experimentally determined temperature rise at locations. The measured temperatures are noted to be accurate to within ~15%. We used the measured Seebeck coefficient of the bulk GaN sample to extract the temperature rise. (**d**) Measured thermoelectric figure of merit ($zT$) for thin and bulk GaN samples, showing steady rise with temperature. An improvement of ~22 to 51% is noted for the thin GaN sample over the bulk GaN sample over the temperature range. **Scale bars of (a), (b) 500 μm, 100 μm.**

Supporting Information

**Tuning Electrical and Thermal Transport in AlGaN/GaN Heterostructures via Buffer Layer Engineering**

*Ananth Saran Yalamarthy[1], Hongyun So[2*], Miguel Muñoz Rojo[3], Ateeq J. Suria[1], Xiaoqing Xu[4], Eric Pop[3,5,6], and Debbie G. Senesky[3,6,7*]*

[1]Department of Mechanical Engineering, Stanford University, Stanford, CA 94305, USA.
[2]Department of Mechanical Engineering, Hanyang University, Seoul 04763, South Korea.
[3]Department of Electrical Engineering, Stanford University, Stanford, CA 94305, USA.
[4]Stanford Nanofabrication Facility, Stanford University, Stanford, CA 94305, USA.
[5]Department of Materials Science and Engineering, Stanford University, Stanford, CA 94305, USA.
[6]Precourt Institute for Energy, Stanford University, Stanford, CA 94305, USA.
[7]Department of Aeronautics and Astronautics, Stanford University, Stanford, CA 94305, USA.

*E-mail: Hongyun So (hyso@hanyang.ac.kr), Debbie G. Senesky (dsenesky@stanford.edu)

Table of Contents:

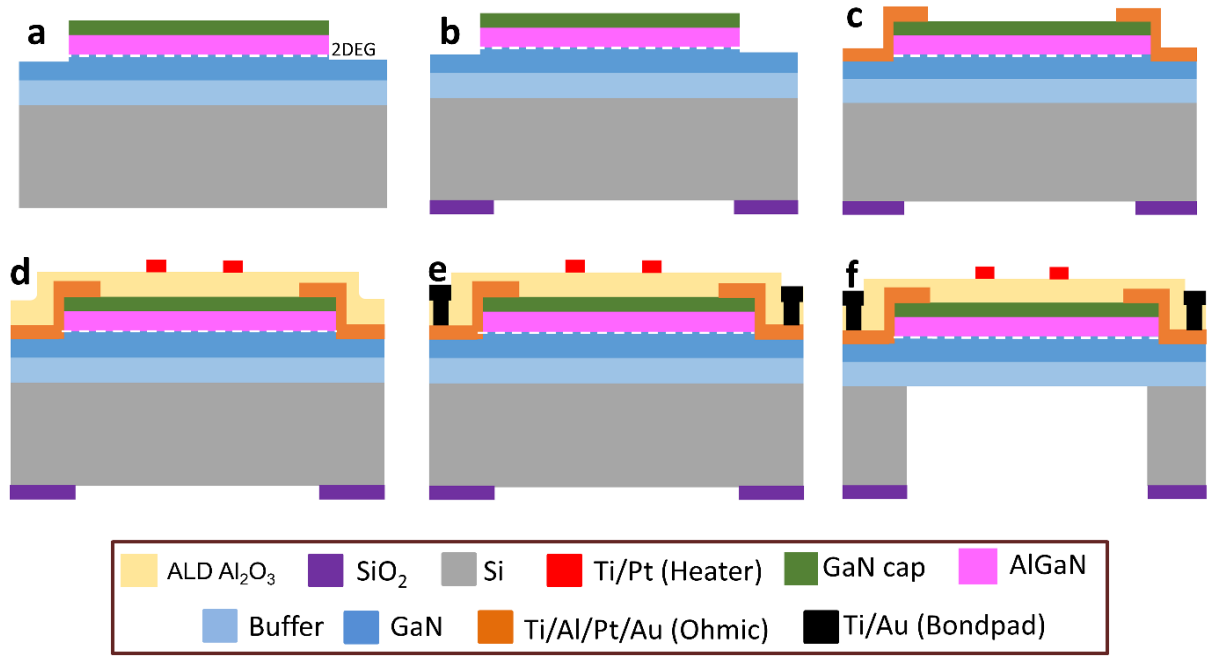

**Figure S1.** Seven-mask process used to fabricate suspended AlGaN/GaN electrical and thermal measurement platform.

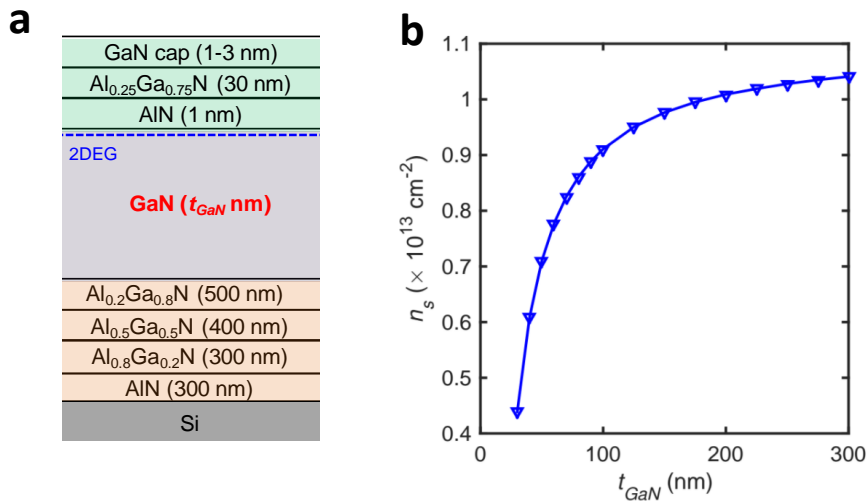

**Figure S2. (a)** Schematic of grown AlGaN/GaN heterostructure, along with the different buffer layers. **(b)** Schrödinger-Poisson simulation of decrease in 2DEG charge density ($n_s$) with reducing the thickness of the GaN buffer layer.

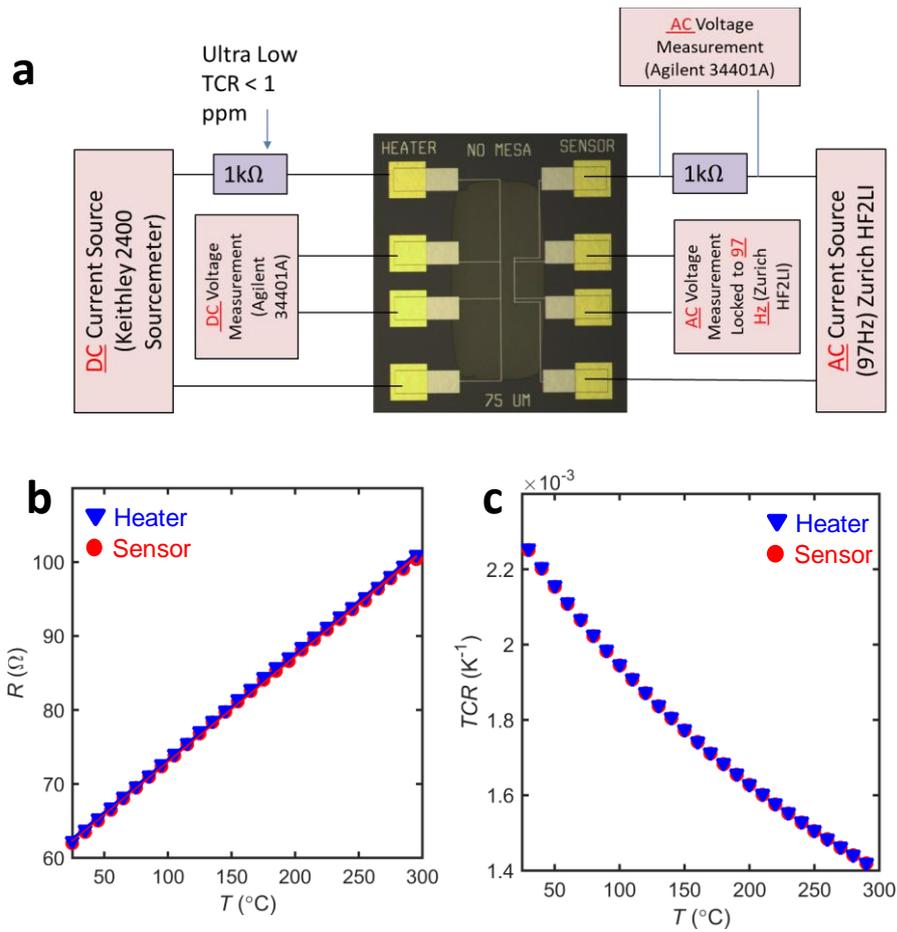

**Figure S3. (a)** Schematic of test setup. We measured the resistance of the heater electrode using a DC multimeter and voltage source with a calibration current of 50 µA. For the sensor side, we used an AC lock-in amplifier with frequency of 97 Hz to minimize self-heating effects. **(b)** Resistance-temperature calibration for the heater and sensor lines showing linear behavior. **(c)** TCR for heater and sensor lines.

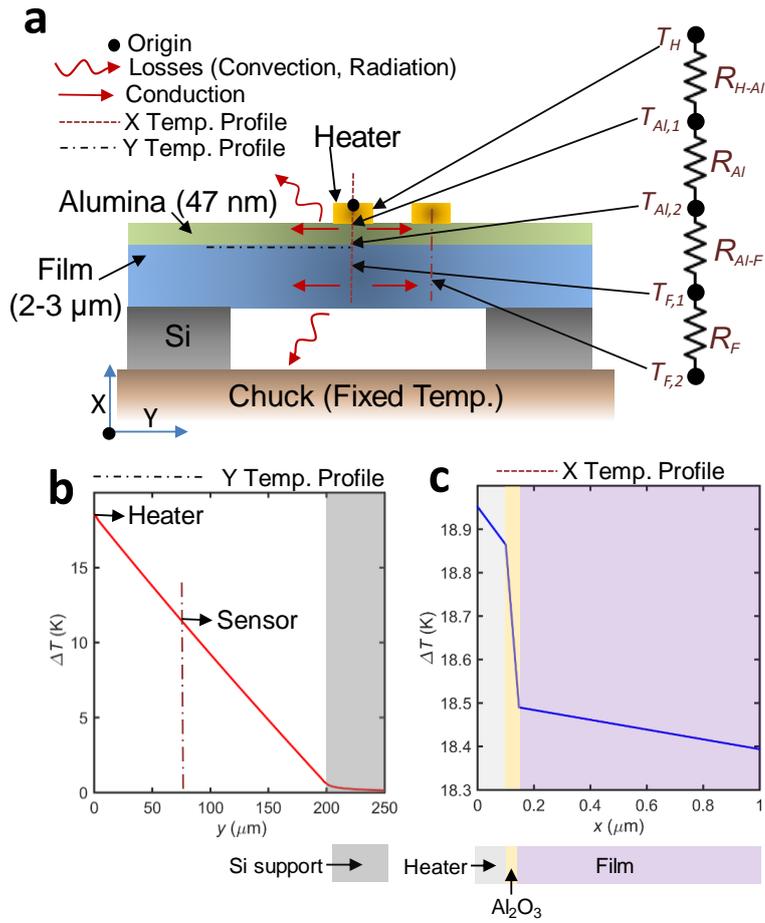

**Figure S4. (a)** Thermal resistance network with the different pathways for heat sinking when current is applied to the heater metal. **(b)** Simulated lateral temperature profile in the AlGaN/GaN/buffer film with ~14 mA heater current. The substrate fixed at 25°C (for the bulk GaN film). **(c)** Simulated vertical temperature drop across the heater/Al$_2$O$_3$/AlGaN/GaN/buffer film for the same conditions in (b).

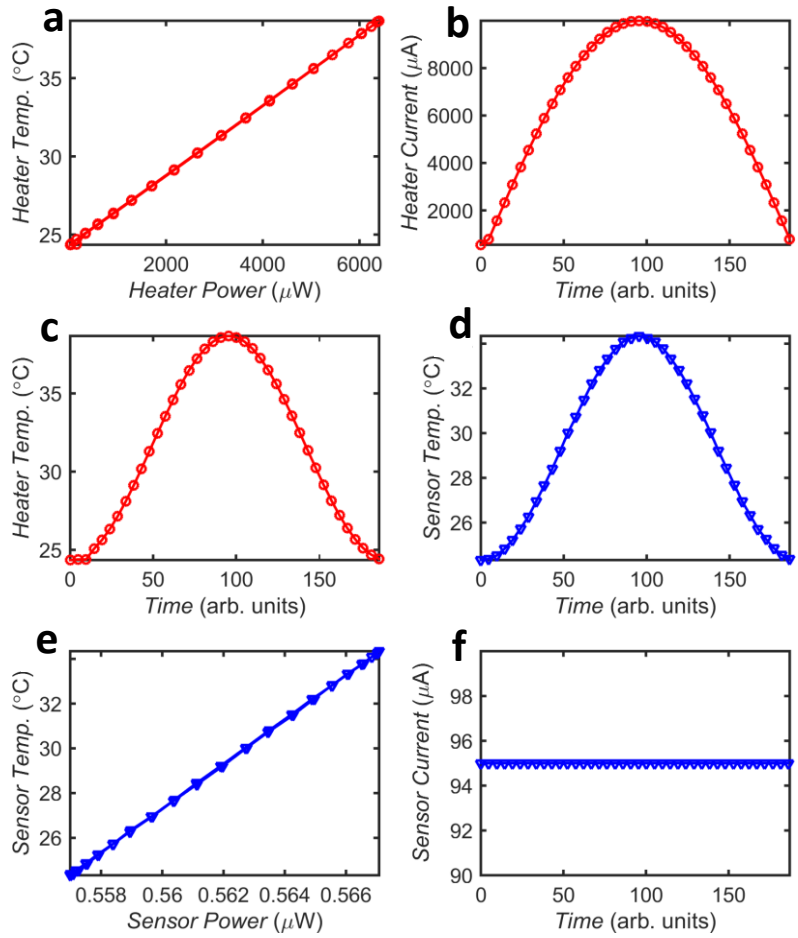

**Figure S5.** Thermal conductivity measurement. Panels **(a)-(c)** are for the heater line, while **(d)-(f)** are for the sensor line. These panels are for the bulk GaN sample with the substrate held at 25°C.

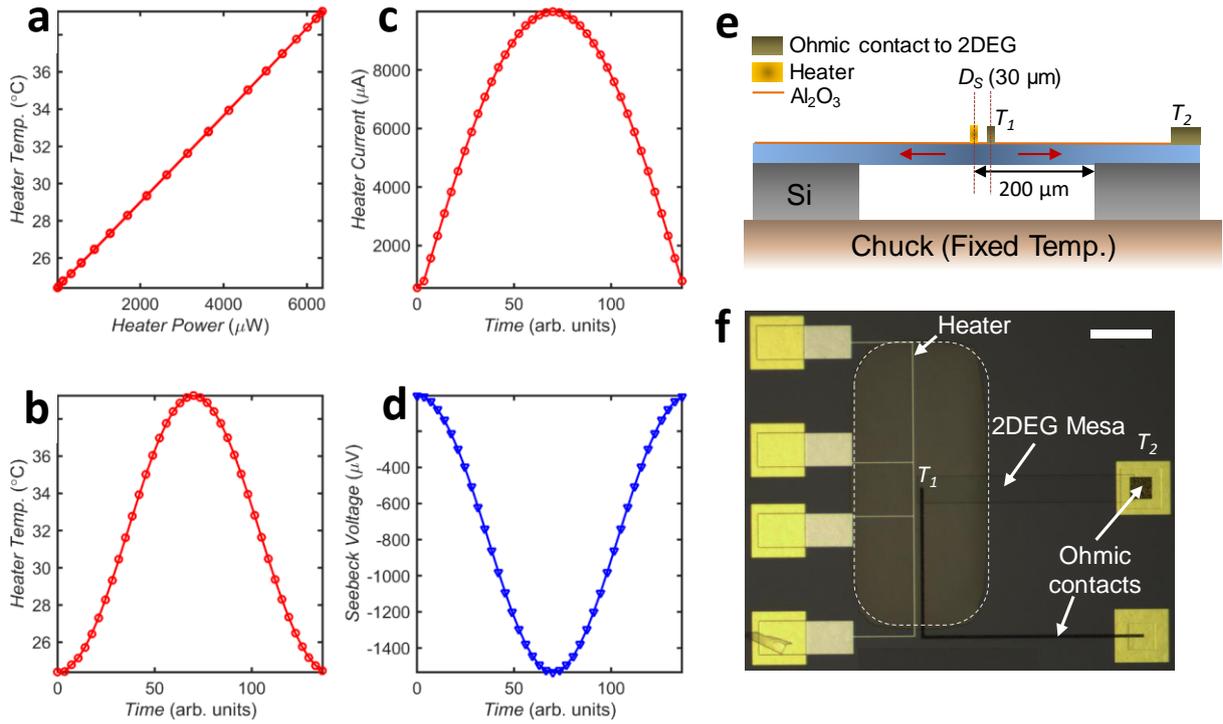

**Figure S6.** Seebeck coefficient measurement. Panels **(a)-(c)** are for the heater line, while **(d)** shows the Seebeck voltage measured in the 2DEG mesa. These panels are for the bulk GaN sample with the substrate held at 25°C. **(e,f)** Cross-section and top view showing the different electrodes for Seebeck coefficient measurement. **Scale bar of (f) 200 μm.**

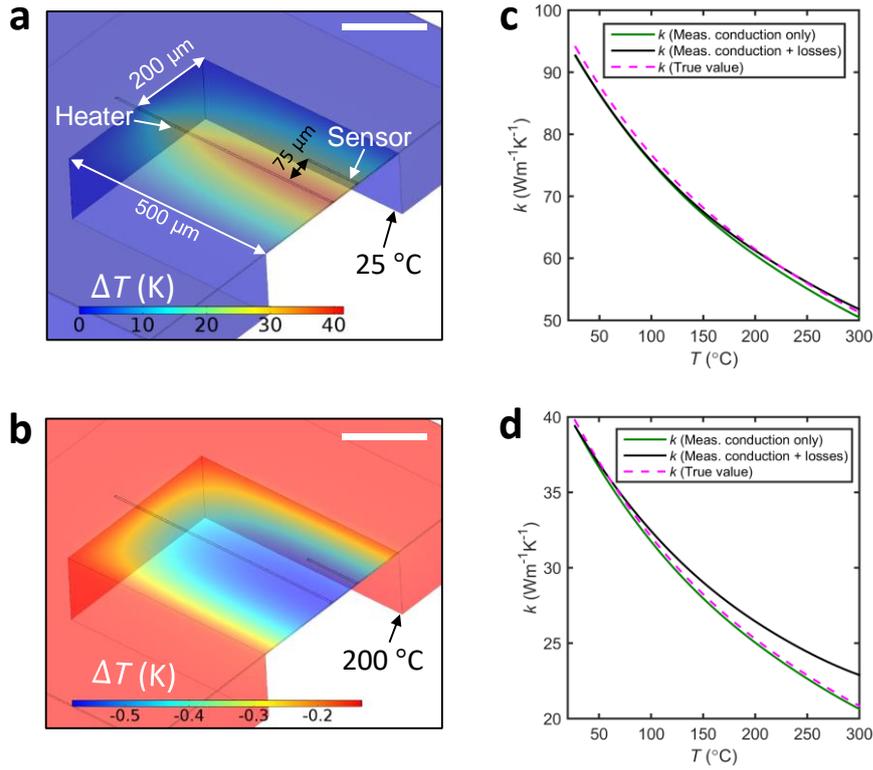

**Figure S7.** (a) Half-symmetric finite-element simulation of experimental structure, showing sample temperature profile when current is applied through the heater with the bottom fixed at room temperature. (b) Temperature profile when no current is applied through the heater with bottom fixed at 200 °C. Notice the cooling in the suspended membrane due to external convection. This effect becomes prominent at temperatures above ~100 °C. (c,d) Thermal conductivity measurements predicted from finite-element model for bulk and thin GaN samples, respectively. At higher temperatures, the measured thermal conductivity is higher than the actual thermal conductivity due to the cooling losses described in Figure S7b. **Scale bars of (a), (b) 200 μm.**

**Table S1.** Estimated thermal conductivity correction factors.

| $T$ (°C) | Bulk GaN | Thin GaN |
|---|---|---|
| 75 °C | -- | 0.055 |
| 100 °C | -- | 0.097 |
| 150 °C | 0.096 | 0.19 |
| 200 °C | 0.17 | 0.31 |
| 250 °C | 0.32 | 0.62 |
| 300 °C | 0.63 | 1.1 |

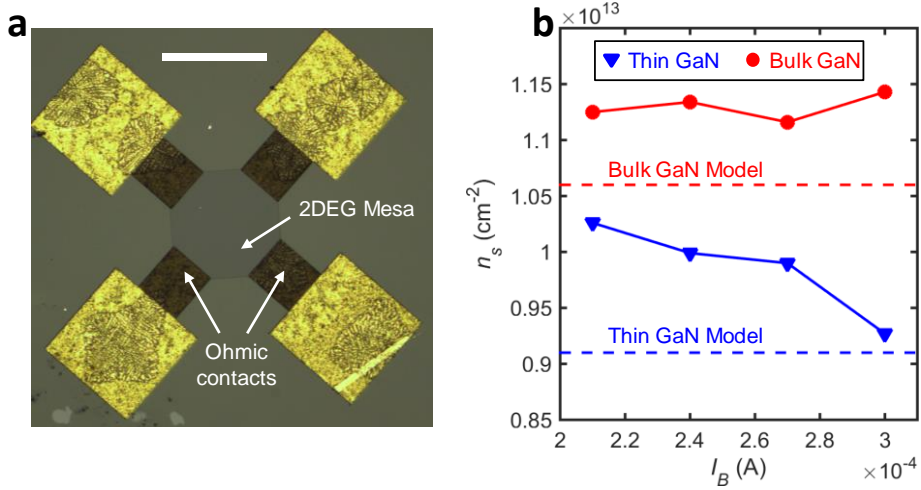

**Figure S8.** Room temperature Hall-effect Measurements. **(a)** Van der Pauw structure used for Hall-effect measurements. An octagonal 2DEG mesa region is used. These devices are co-fabricated with the other electrical and thermal transport test structures described in this article. **(b)** Extracted sheet density ($n_s$) for a range of applied bias currents ($I_B$) at room temperature. The estimates from the Schrödinger-Poisson model are shown using dashed lines. **Scale bar of (a) 200 μm.**

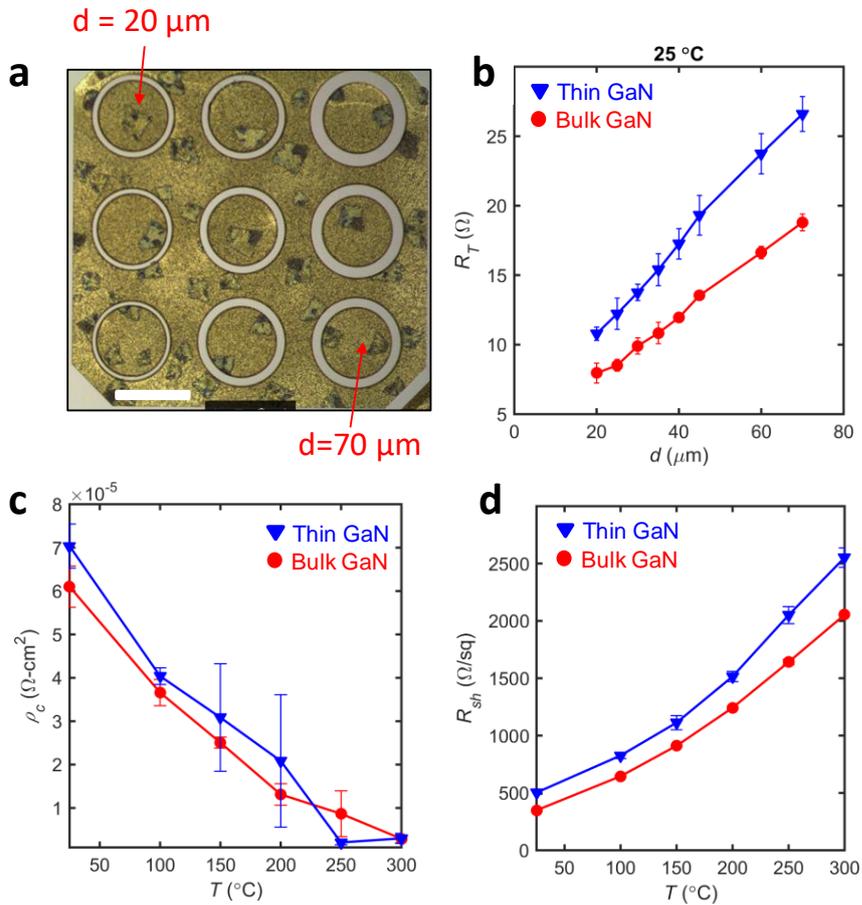

**Figure S9.** CTLM Measurements. **(a)** CTLM test array for measurement of contact & sheet resistance with gap spacing varying from 20 μm to 70 μm. **(b)** Transfer resistance versus gap spacing for the thin and bulk GaN samples to extract sheet & contact resistance at 25°C. **(c)** Temperature dependent contact resistance for the thin and bulk GaN samples. **(d)** Temperature dependent sheet resistance for the thin and bulk GaN samples. Error bars are obtained from measurements across 4 identical CTLM test arrays. **Scale bar of (a) 500 μm.**

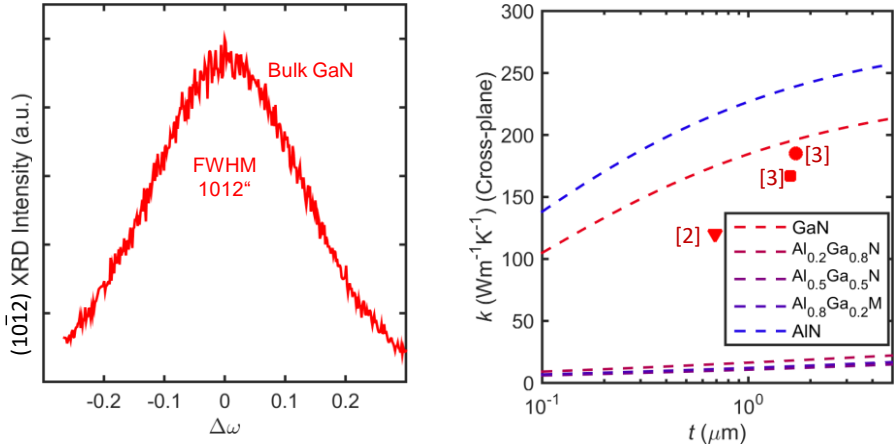

**Figure S10.** **(a)** XRD scan to estimate the density of edge-type dislocations in the bulk GaN sample. The edge type dislocation density is estimated as $\approx 3\times10^9$ cm$^{-2}$ following the methods described in Lee et al.[1] **(b)** Cross-plane thermal conductivity estimates of the layers in our composite film as a function of thickness. The experimental data points correspond to cross-plane measurements in GaN.[2,3]

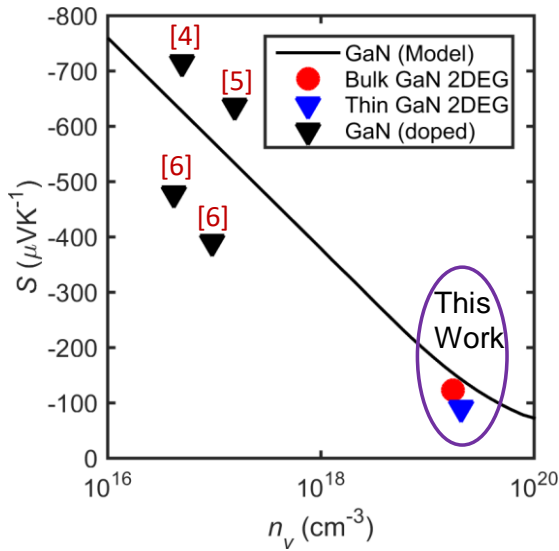

**Figure S11.** Room temperature Seebeck coefficients ($S$) as a function of carrier concentration ($n_v$) in GaN. The black line represents the Seebeck coefficient model. The experimental data points[4–6] and the model are at room temperature. All the data points correspond to doped, bulk films. The measured values for the 2DEG in the bulk and thin GaN sample in this work are also plotted.

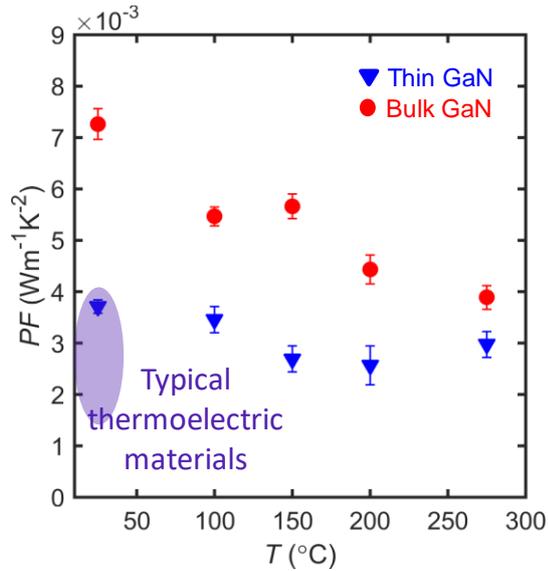

**Figure S12.** Power factor of the 2DEG in the thin and bulk GaN samples. The power factor for state-of-the-art thermoelectric materials is usually between $1.5\times10^{-3}$ and $4\times10^{-3}$ $Wm^{-1}K^{-2}$ at room temperature.[7]

**Supplementary Note 1:** Fabrication Process

Figure S1 shows the seven-mask process to fabricate the fully-suspended AlGaN/GaN platform for thermal measurements. A schematic of the heterostructure showing the different buffer layers and the silicon substrate is illustrated in Figure S2a. The AlGaN/GaN/buffer heterostructure for the thin and bulk GaN samples was grown using an in-house metal organic chemical vapor deposition (MOCVD) chamber on a Si (111) substrate. In order to define the 2DEG mesa, we etched the AlGaN/GaN layers to a depth of ~100 nm using an inductive coupled plasma technique with $BCl_3/Cl_2$ gases as shown in Figure S1a. This was followed by the deposition of ~4 μm PECVD oxide on the backside and selectively patterned to define the Si removal region, as depicted in Figure S1b. The Ohmic contacts to the 2DEG were patterned by depositing Ti/Al/Pt/Au (20/100/40/80 nm) followed by a rapid thermal anneal (RTA) in $N_2$ ambient at 850°C for 35 seconds (Figure S1c). Next, we deposited ~47 nm of atomic-layer deposited (ALD) $Al_2O_3$ followed by patterning Ti/Pt (10/100 nm) heater and sensor metal lines, as shown in Figure S1d. To deposit Ti/Au (20/200 nm) bondpad metal, we opened vias in the ALD film using a 20:1 buffered oxide etch for ~2 min (Figure S1e). In order to release the AlGaN/GaN/buffer heterostructure, Si was finally etched from the backside using a deep reactive ion etching (DRIE) technique, stopping at the buffer/Si interface. X-Ray diffraction data for the AlGaN/GaN/buffer layers in available in our former work.[8] After suspension, the total thickness of the heterostructure layers was obtained as ~1.695 μm for the thin GaN heterostructure and ~2.85 μm for the bulk GaN heterostructure from the SEM cross-section images.

**Choice of buffer layer thicknesses:**

The thermal conductivity of the AlGaN/GaN/buffer hetero-structure is typically determined by the thermal conductivity of the GaN buffer layer. The thickness of the GaN ($t_{GaN}$) layer in the thin GaN sample was designed to lower the thermal conductivity of the buffer layer structure due to size effect, while preserving the charge density $n_s$ of the 2DEG. This effect is shown in Figure S2b, which depicts the rapid decline in $n_s$ with $t_{GaN}$ due to band bending in the AlGaN and GaN layers from the Schrödinger Poisson model. As the GaN layer thickness decreases below ~100 nm, the decline is much sharper as the strain difference in the AlGaN/GaN layers decreases, leading to the GaN layer becoming pseudomorphic with the buffer layers beneath.

**Supplementary Note 2:** Test Setup and Calibration

Figure S3a shows the test setup used to measure the in-plane thermal conductivity of the AlGaN/GaN hetero-structure. In order to ensure accuracy in the thermal conductivity measurements, we performed careful resistance versus temperature calibration for the Ti/Pt heater and sensor lines. For the heater line, a DC current source (Keithley 2400) and a DC voltage source (Agilent 34401) were used to measure the resistance of the Ti/Pt trace. To measure the resistance of the sensor line, we used an AC voltage lock-in amplifier (Zurich Instruments HF2LI) with a lock-in frequency of 97 Hz. AC voltage measurement across a fixed resistor (1 kΩ, ultra-low TCR of less than 1 ppm) was used to infer the AC current from the applied AC voltage. The lock in-amplifier was chosen for the sensor side to minimize self-heating effects and block environmental noise. In order to calibrate the resistances of both lines, the substrate of the suspended membrane was attached to a temperature controlled chuck using high vacuum thermal grease (Apiezon, Inc.). A current value of ~100 μA was carefully chosen for the purpose of resistance calibration to avoid self-heating effects.

Figure S3b and Figure S3c show the calibration curves of resistance ($R$) and temperature coefficient of resistance (TCR) till 300°C. Initially, we obtained a non-linear resistance-temperature calibration curve due to the effects of alloying in the Ti/Pt metal. In order to alleviate this problem, the entire sample was heated to ~400°C and held for ~10 minutes to anneal the Ti/Pt metals. Upon annealing, the resistance calibration curve is found to be extremely linear as can be seen in Figure S3b ($R^2 \approx 0.998$), and this is later used to extract the temperature of the heater line when heating power is applied to it to extract the thermal conductivity of the AlGaN/GaN heterostructure. The fitted slope is ~0.148 $\Omega K^{-1}$, as can be seen in Figure S3b. Note that the plotted resistance values are obtained by averaging over 20 measurements spaced by 2 seconds at each substrate temperature. In each case, the error bar (defined as the range) for the resistance measurement is smaller than the size of the markers in Figure S3b and Figure S3c, with steadily increasing error bar as the temperature increases. Typical values for the error bars are ~1.5 m$\Omega$ at 25°C with a steady increase to ~50 m$\Omega$ at 300°C. Using the fitted slope, this implies an error of $< +/- 0.5$°C for the measured temperature even at 300°C, which is accounted for in the extraction of thermal conductivity. A similar calibration procedure was performed for the heater line in the Seebeck coefficient measurement platform for the bulk and thin GaN samples.

**Supplementary Note 3:** Finite-Element Model, Measurement Process and Error Correction

In order to understand the errors associated with the measurement and extract the thermal properties of the AlGaN/GaN/buffer hetero-structure, a 3-D finite-element model using a commercial software (COMSOL) was implemented. Figure S7a shows a half-symmetric finite-element simulation of experimental structure, showing sample temperature profile when current is applied through the heater with the substrate fixed 25°C. The heater and sensor lines have a width ($W$) of 5 μm, and are spaced by a distance ($D_{HS}$) of 75 μm. The location of the heater and sensor resistances on the suspended membrane ($R_H$ and $R_S$) are chosen such that the heat transfer can be well approximated as 1-D,[9] which facilitates the extraction of the thermal conductivity. The typical variation of temperature along the length of these resistors is estimated to be < 0.01% from the simulation model, which supports this assumption. Figure S4a shows a cross-section schematic of the thermal resistance network with the different pathways for heat sinking when current is applied to the heater metal. In the absence of external convective and radiative losses and negligible contribution of the Alumina film to the in-plane heat conduction, the thermal resistance of the film ($R_F$) can be estimated as:

$$R_F = \frac{2(T_H - T_S)}{P_H} - 2R_{Al} - \frac{2(R_{mox} + R_{oxg})}{A_H} \tag{S1}$$

where $T_H$ and $T_S$ are the heater and sensor line temperatures, $P_H$ is the input power to the heater and $R_{Al}$ is the thermal resistance of the Al$_2$O$_3$ layer, $A_H$ is the area projected area of the heater electrode (5 μm × 200 μm), $R_{mox}$ is the thermal boundary resistance of the Heater/Al$_2$O$_3$ interface and $R_{oxg}$ is the thermal boundary resistance of the Al$_2$O$_3$/GaN interface. The thermal conductivity of the film can be extracted from $R_F$ and the known film dimensions. The simulated temperature drop from the heater to the sensor in the GaN film is linear, as can be seen along the Y direction in Figure S4b. Equation S1 suggests that the thermal conductivity of the film can be measured accurately in the limit of $R_F \gg R_{Al+Interfaces}$. The resistance ratio is analytically estimated as:

$$\frac{R_{Al+Interfaces}}{R_F} = \frac{T_{Al} k_F T_F}{k_{Al} W D_{HS}} + \frac{(R_{mox} + R_{oxg}) k_F T_F}{W D_{HS}} \tag{S2}$$

Where $T_{Al}$ and $T_F$ are the thicknesses of the alumina and AlGaN/GaN/buffer film, respectively. We used a thermal boundary resistance of $2.8 \times 10^{-8}$ m$^2$K W$^{-1}$ for $R_{mox}$.[10] Although data for the thermal boundary resistance across the Al$_2$O$_3$/GaN film interface is not available, we estimated $R_{oxg}$ ~ $1 \times 10^{-8}$ m$^2$KW$^{-1}$ based on measurements across amorphous dielectric/Si interfaces,[11] since GaN and Si have similar Debye temperatures.[12,13] At room temperature for the bulk GaN film, using $T_F \approx 3$ μm, $T_{Al} \approx 47$ nm, $k_{Al} \approx 2$ Wm$^{-1}$K$^{-1}$[10] and $k_F \approx 115$ Wm$^{-1}$K$^{-1}$, we estimate a thermal resistance ratio of ~5.7%. For the thin GaN film, since $T_F \approx 2$ μm and $k_F \approx 45$ Wm$^{-1}$K$^{-1}$, the thermal resistance ratio is ~1.48%, thus the error due to loss in Al$_2$O$_3$ is smaller than the bulk GaN film. Note that these values also support the assumption in Equation S1 that Alumina does not contribute to the in-plane heat conduction, since $k_{Al}T_{Al} \ll k_F T_F$. At higher temperatures, the error due to this effect is less pronounced as $k_{Al}$ is found to increase,[10] while $k_F$ further decreases, as can be seen in Figure 3b. The loss via the Al$_2$O$_3$ and the interfaces can also be observed via the COMSOL model, as can be seen in Figure S4c (bulk GaN film, substrate at 25°C). In the model, in addition to using $k_{Al} \approx 2$ Wm$^{-1}$K$^{-1}$, The temperature drops by ~0.47 K in Al$_2$O$_3$ and interfaces on either side relative to ~7.4 K in the GaN film (across $D_{HS}$), giving an error of ~6.35%.

Figure S5 shows the typical thermal conductivity measurement procedure for our films. These plots are from experiments with the bulk GaN sample. In this experiment, the substrate is held at 25°C. The sensor is maintained at the calibration current of ~100 μA (Figure S5f), while the heater current is ramped up in a half-sinusoid from its initial calibration value (Figure S5b). Before each resistance measurement, we wait for 2 seconds after the current ramp to allow the system to equilibrate. The waiting interval of 2 seconds was chosen based on an estimation of a thermal time constant of ~2 milli-seconds for the suspended membrane from COMSOL simulations. The heater & sensor temperature (converted from the resistance via the calibration curve in Figure S3b) track the current pattern, with the initial temperature equal to the substrate temperature, as seen in Figure S5c and Figure S5d. The extracted temperature difference can be used to calculate the in-plane film thermal conductivity via Equation S1, after accounting for the Al$_2$O$_3$ temperature drop and external losses, which are discussed later. In addition, we also ensured that hysteresis did not occur in our heater and sensor lines. This is clear from observing the temperature versus power lines in Figure S5a and Figure S5e.

Figure S6 shows a typical Seebeck coefficient measurement procedure. Similar to the thermal conductivity measurement, the heater current is ramped up from its calibration value, setting up a lateral temperature gradient along the 2DEG mesa which translates to a measurable Seebeck voltage (Figure S6d). The measured Seebeck voltage includes a minor contribution from the temperature drop across the Ti/Al/Pt/Au Ohmic contacts to the 2DEG (visualized in Figure S6f). At room temperature, we measured the Seebeck voltage across the 2DEG mesa and the Ohmic metal line for an identical temperature gradient. Using this, we estimated the contribution of the Ohmic metal line to be less than 2% of the overall Seebeck voltage, and thus neglected its effect in subsequent measurements. The Seebeck voltage of the 2DEG is given as $S_{2DEG}=V_{2DEG}/(T_1-T_2)$, as depicted in Figure S6e and Figure S6f. The temperature at the contact outside the suspended region ($T_2$) is assumed to be at the substrate temperature. The temperature drop in the silicon supported region is <1% of the total temperature drop ($T_1$-$T_2$) (Figure S4b), thus, the contribution to the Seebeck coefficient from the supported region can be ignored. $T_1$ is related to the heater temperature $T_H$ as:

$$\frac{(T_H - T_1)}{P_H} = R_{Al} + \frac{R_F}{2} + \frac{(R_{mox} + R_{oxg})}{A_H} \tag{S3}$$

where $R_F$ is calculated using the measured film thermal conductivity and a length of 30 $\mu$m ($D_S$, depicted in Figure S6e) and $R_{Al}$ is calculated as discussed earlier.

Finally, external losses from convection are significant at high temperatures which lead to errors in the thermal conductivity measurement. This can be seen in Figure S7b, which shows the temperature profile when no current is applied through the heater with bottom fixed at 200°C for the bulk GaN film. Notice the cooling in the suspended membrane due to external convection, leading a relative temperature difference between the heater and the sensor line. This underestimates the $\Delta T$ between the heater and the sensor line, leading a higher measured thermal conductivity than the true value. The actual temperature gradient can be expressed as $\Delta T_{real} = \Delta T_{meas} + \Delta T_{corr}$, where $\Delta T_{corr}$ is the temperature gradient from the heater to the sensor with no current applied. The prominence of this effect is seen in Figure S7c and Figure S7d, which shows the simulated true and estimated measurements of the thermal conductivity for the normal GaN and thin GaN film. The error in the measured thermal conductivity is estimated to be ~4% and ~10% for the normal and thin GaN films from simulations, using an external convection coefficient of 10 Wm$^{-2}$K$^{-1}$. The correction factor $\Delta T_{corr}$ can be estimated from the natural convection coefficient ($h$). At progressively higher temperatures, a non-zero Seebeck voltage is observed when no current is applied in the heater line due to the effect shown in Figure S7b. Then, using the most recently corrected thermal conductivity value, we estimated the value of $h$ required to produce the observed non-zero Seebeck voltage using our knowledge of the Seebeck coefficient from the COMSOL model. Following this, we obtained the correction factor $\Delta T_{corr}$ for the current thermal conductivity value. Typical values for $h$ estimated using this procedure are in the range of 12-14 Wm$^{-2}$K$^{-1}$, which are reasonable coefficients for natural convection. The correction factors ($\Delta T_{corr}$) we obtained for the normal GaN and thin GaN films for the different substrate are tabulated in Table S1.

**Supplementary Note 4:** Schrödinger-Poisson Model Notes and Validation

The models for the bulk and thin GaN heterostructures were made using a commercially available Schrödinger-Poisson device physics simulator (NextNano Inc.).[14] In both models, we set the barrier height for the GaN capping layer to 1 eV, based on the assumption that the surface is exposed to air.[15] The entire structure is simulated using a 1-D grid size of 0.5 nm, except in the region where the 2DEG quantum well forms, where we used a finer grid of 0.1 nm. In both cases, the simulation requires a substrate to determine the strain in the heterostructure. For the bulk GaN model, we used a thick 500 nm GaN layer as the substrate to converge the 2DEG sheet density ($n_s$). However, for the thin GaN model, the GaN layer was set to 100 nm, while the thick layer below it, $Al_{0.2}Ga_{0.8}N$, was used as the substrate to determine the strain level. A 500 nm thick $Al_{0.2}Ga_{0.8}N$ layer was sufficient to converge the charge density in the thin GaN model. A lattice temperature of 300 K was used for the both models. The source code for both models is available at https://github.com/ananthy/GaNThick.

The Schrödinger-Poisson model is validated by comparing the sheet densities obtained from the model with the values obtained from Hall-effect measurements. We first recall that we obtained a sheet density $n_s = 1.06 \times 10^{13}$ cm$^{-2}$ and $n_s = 0.91 \times 10^{13}$ cm$^{-2}$ for the bulk and thin GaN heterostructures from the simulations, respectively. We recall that the sheet density is ~16% lower for the thin GaN sample due the reduced 2DEG quantum well depth ($t_{2D}$), but has a similar peak volumetric charge density as the bulk GaN sample, as discussed in Figure 3. A simple, 4-contact Van der Pauw structure was used to characterize the sheet density via Hall-effect measurements. The structure is shown in Figure S8a. The conducting 2DEG area for each Hall-effect plate was 200 μm × 200 μm. Using a range of bias current levels ($I_B$), an external magnetic field ($B$) of ~1 mT, $n_s$ is related to the Hall voltage $V_H$ as

$$n_s = \frac{I_B B}{q V_H} \qquad (S5)$$

where $q$ is the electronic charge. The results from these experiments performed at room temperature are depicted in Figure S8b. Good match between the sheet densities obtained from the experiments and the model is observed. The values from the experiments are about 5-10% higher than those predicted from the Schrödinger-Poisson model, however, we note that the average difference in the sheet densities of the thin and bulk GaN samples from experiments (~14.3%) and the model (~16.5%) is predicted accurately, which serves as a validation for the model, in particular, the thin GaN heterostructure. The fluctuation in the experimentally observed $n_s$ values for different bias currents ($I_B$) could arise from non-linearity in the current-voltage characteristics and other sources such as offset voltage from thermal effects.[16]

**Supplementary Note 5:** Circular Transfer Length Method (CTLM)

CTLM test structures with gap spacing ($d$) varying from 20-70 μm (Figure S9a) were used to measure the sheet resistance of the 2DEG and the contact resistance for the thin and bulk GaN samples. The inner radius ($L$) of the circular test structures was designed to be 500 μm, such that the approximation $L \gg d$ is valid. In the regime where $L \gg d$, the total resistance $R_T$ between the inner and outer contact can be expressed as[17]:

$$R_T = \frac{R_{sh}(d + 2L_T)}{2\pi d}\log\left(1 + \frac{d}{L}\right) \quad (S4)$$

where $L_T$ is the transfer length and $R_{sh}$ is the sheet resistance of the 2DEG. Here, $L_T$ is related to $R_{sh}$ and the contact resistance $\rho_c$ as:

$$L_T = \sqrt{\rho_c/R_{sh}} \quad (S5)$$

Figure S9b illustrates the variation of the total resistance $R_T$ with gap spacing $d$ (Equation S4) for the thin and bulk GaN samples at 25°C, which is used to extract $R_{sh}$ and $\rho_c$. Using the 2DEG thickness, $t_{2D}$, the extracted $R_{sh}$ is used to extract the 2DEG conductivity, as plotted in Figure 3c from 25 °C to 300°C. The values of $R_{sh}$ for the bulk and thin GaN sample are also shown in Figure S9d. The contact resistance values are noted to be in the right range for typical 2DEG Ohmic contacts[18] ($10^{-5}$–$10^{-6}$ Ω-cm$^{-2}$) and decreasing with increasing temperature due to enhanced thermionic field emission across the GaN/AlGaN layers,[19,20] as observed in Figure S9c. A similar magnitude of decrease has also been observed in former work.[19]